\documentclass[prc,preprint,aps,showpacs,superscriptaddress]{revtex4}

\usepackage{graphicx}
\usepackage{dcolumn}
\usepackage{bm}

\begin{document}

\preprint{Vetter 2005 O-14}

\title{Half-life of $^{14}$O}

\author{J.T. Burke}
\altaffiliation[Present Address:  ]{Lawrence Livermore National Laboratory, Livermore, CA 94550}
\affiliation{Department of Physics, University of California, Berkeley;
and Nuclear Science Division,
Ernest Orlando Lawrence Berkeley National Laboratory, Berkeley, CA 94720}
\author{P.A. Vetter}
\affiliation{Nuclear Science Division, Ernest Orlando Lawrence Berkeley National Laboratory, 
Berkeley, CA 94720}
\author{S.J. Freedman}
\affiliation{Department of Physics, University of California, Berkeley;
and Nuclear Science Division,
Ernest Orlando Lawrence Berkeley National Laboratory, Berkeley, CA 94720}
\author{B.K. Fujikawa}
\affiliation{Nuclear Science Division, Ernest Orlando Lawrence Berkeley National Laboratory, 
Berkeley, CA 94720}
\author{W.T. Winter}
\affiliation{Department of Physics, University of California, Berkeley;
and Nuclear Science Division,
Ernest Orlando Lawrence Berkeley National Laboratory, Berkeley, CA 94720}

\date{\today}

\begin{abstract}
We have measured the half-life of $^{14}$O, a superallowed $\left( 0^{+}
\rightarrow 0^{+}\right)$ $\beta$ decay isotope.  The
$^{14}$O was produced by the $^{12}$C($^{3}$He,n)$^{14}$O reaction
using a carbon aerogel target.  A low-energy ion beam of $^{14}$O was
mass separated and implanted in a thin beryllium foil. The beta
particles were counted with plastic scintillator detectors.  We find
$t_{1/2} = 70.696\pm 0.052$~s.  This result is $1.5\sigma$ higher than
an average value from six earlier experiments, but agrees more closely
with the most recent previous measurement.
\end{abstract}


\maketitle

Superallowed nuclear beta decays can be used to determine the
effective weak vector-coupling constant ($G_{V}$) for the nucleon.
The Cabbibo-Kobayashi-Maskawa (CKM) matrix element $V_{ud}$ is
obtained from $G_{V}$ and the corresponding muon decay constant,
$G_{\mu}$, after appropriate radiative corrections are applied.
Uncertainties in $V_{ud}$ presently contribute the largest uncertainty
to a precision unitarity test of the matrix.  To determine the ${\cal
F}t$ values to the required precision, the half-lives, branching
ratios, and Q-values of superallowed transitions must be measured
precisely.  Several small but important radiative corrections must be
determined reliably \cite{towner02}.  The theoretical uncertainty in
$V_{ud}$ is currently thought to be dominated by nuclear structure
dependent corrections.  To attempt to assess the reliability of the
theoretical corrections, some new work is directed to measuring
superallowed transitions in higher-Z nuclei where the corrections are
larger \cite{hardy03}.  However, it remains important to reduce
experimental uncertainties in the low-Z systems where the
corrections are small.  Radioactive beam techniques provide a new
opportunity to study superallowed $\beta$ decays by making accessible
systems with higher Z or $T_{z} = -1$, which lie farther from
stability.  These techniques can also be used to study low-Z
superallowed beta decays with better precision by using very pure,
mass-separated samples, as in Ref.\ ~\cite{gaelens01}.

We have measured the $^{14}$O half-life using a mass-separated
radioactive ion beam, providing a relatively pure sample.  With a
low contamination sample, we could use simple fast organic
scintillators to detect the beta particles emitted by $^{14}$O,
instead of the gamma rays as in most previous experiments.  Organic
scintillators have small, well-controlled dead time as compared to the
much slower germanium detectors used in previous experiments such as
Ref.\ ~\cite{clark73}.  The authors of Ref.\ ~\cite{barker04}
identified a serious and unappreciated systematic effect in germanium
detectors used for high-precision lifetime measurements.

\begin{figure}
\includegraphics[width=0.5\textwidth]{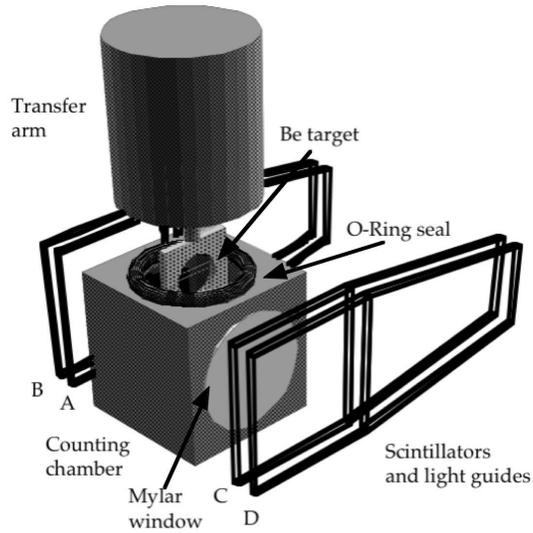}
\caption{\label{fig:chamber}
The arrangement of the detectors, counting chamber, and target foil
positioned for decay data acquisition.  The vacuum chamber enclosing
the transfer arm is not shown, nor are the photomultiplier tubes and
wrapping on the scintillators and light guides.}
\end{figure}

The $^{14}$O in this experiment was produced via the
$^{12}$C($^3$He,n)$^{14}$O reaction using a 20~MeV $^{3}$He beam (up
to 10 p$\mu$A) from the 88-Inch cyclotron at Lawrence Berkeley
National Laboratory.  The target was a piece of carbon aerogel
(0.25~g/cm$^3$) heated to 2000~K by the cyclotron beam and 200 W of
additional electrical power from resistive heating.  The $^{14}$O
generated in the target evolved as gaseous CO and CO$_{2}$.
Radioactive gas diffusing from the target was pumped through a ten
meter vacuum line by a magnetically levitated turbomolecular pump into
an electron cyclotron resonance (ECR) ion source, described in Ref.\
~\cite{wutte98}.  The turbopump exhaust line was run through a cold
trap (at -78 C) at the inlet to the ion source to remove condensable
contaminant gases, improving the ion source performance.  We estimated
the fraction of the $^{14}$O released from the target by venting the
exhaust from the turbopump into a closed vacuum chamber and counting
the characteristic $2.3$ MeV gamma ray from $^{14}$O decay in this
volume.  The $^{14}$O yield from the target was calculated from the
cross-section\cite{osgood64}.  We conclude that $45 \pm 5 \%$ of the
$^{14}$O was released from the target at 2000~K.  The ion source
produced a 54~keV, mass-separated $^{14}$O$^{2+}$ ion beam which was
implanted into a 150~$\mu$m thick beryllium foil (thin enough for the
$\beta$ particles to pass through to trigger the detectors on both
sides of the foil).  The mean ion implantation depth was estimated to
be 136~nm using an ion stopping and range calculation software
package.  At this depth and at the slightly elevated temperature of
the foil (caused by ion beam heating), the diffusion of the $^{14}$O
activity out of the foil during the count cycle is negligible.  The
target foil was shuttled 82~cm in vacuum in 10~s to a separate,
shielded observation chamber with a magnetically coupled manipulator.
The observation chamber was a hollow, aluminum cube (2.5~cm) with two
50~$\mu$m thick aluminized mylar windows on opposite sides.  A buna
rubber O-ring sealed the manipulator arm, preventing any diffusion of
radioactive gases into or out of the chamber during the counting
cycle.  The 0.1~cm$\times$2.54~cm$\times$2.54~cm plastic scintillators
were located 4~mm from the windows.  Tapered light guides coupled each
of the four scintillators to a Hamamatsu R647 photomultiplier tube.
Two pairs of scintillators were placed on opposite sides of the cube,
with one scintillator 3~mm in front of another. The arrangement of the
detectors around the counting chamber is shown in
Figure~\ref{fig:chamber}.  Beta particles from the source passed
through the mylar windows, producing minimum ionizing signals in the
scintillation detectors. The discriminator thresholds were set below
the most probable minimum ionizing energy peak to mitigate the effects
of gain shifts of the detectors.

Twenty-nine counting runs were performed during a 30 hour period.  The
target foil was exposed to the $^{14}$O$^{2+}$ ion beam for 200~s.
The source activity reached approximately $10^{6}$ decays per second
during the bombardment.  Following this exposure, a gate valve
interrupted the ion beam, and the target foil was shuttled to the
observation chamber. Scalers recorded the rate of three beta detectors
(a fourth counter failed during the run and was not included in the
analysis) for a period of 4000~s. At the end of the observation
period, the foil was returned to the beam line, the gate valve was
opened, and the procedure repeated.  The 29 data sets each contained
8000 time bins spaced at 0.5~s intervals.  The counting scalers were
gated off for the last 500~$\mu$s of each time bin to allow the data
to be read out.  We used a very long observation time (56 half-lives)
in each run to precisely determine the background rate and to search
for radioactive contaminant species.  By design, the dead time of the
system was dominated by the width of the final logic pulse sent to the
scalers. This provided a single, well-characterized dead time much
longer than other dead time contributions earlier in the electronic
logic chain.  The logic signal from each detector was sent to three
separate scaler channels with different nominal dead times: 400~ns
non-extendable, 700~ns extendable, and 400~ns extendable. This allowed
analysis with different dead times for the same data set.  The minimum
dead time was long enough to avoid effects of PMT after pulsing,
detected in about $10^{-4}$ of the PMT pulses.  The width of the count
pulse sent to the scaler (the largest single component of the dead
time) was measured off-line using a calibrated time-to-digital
converter.  The total dead time of the system was also checked using a
radioactive source to determine the fractional loss in the rate in the
final scalers compared to a fast analog scaler measuring the PMT
signals.  These two methods gave dead times differing by less than
9~ns, which we take to be the dead time uncertainty.

\begin{figure}
\includegraphics[width=0.5\textwidth]{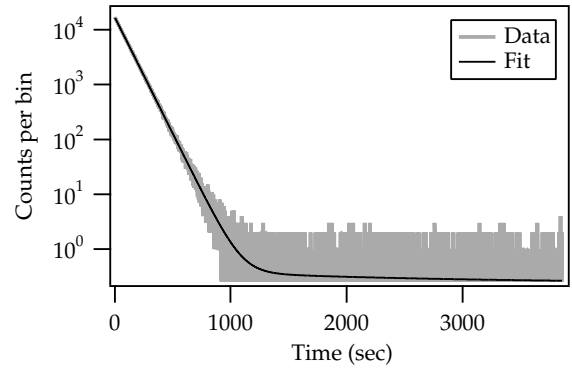}
\caption{\label{fig:decaycycle} 
Decay data and the fit curve for a single run for one detector (C) in
one dead time channel (400 ns, non-extending).}
\end{figure}

The data were analyzed by fitting to exponential decay curves and a
flat background.  The fits used maximum likelihood curve-fitting
rather than chi-squared minimization, based on the arguments in Ref.\
~\cite{baker84}, which argues that chi-squared minimization is
unsuitable for data spanning a wide range of statistical uncertainty
or when the number of data counts per bin becomes small.  The free
parameters in the fit function were the initial decay rates for
$^{14}$O and potential contaminants, a constant background term, and
the half-life of $^{14}$O. The dead times were fixed in the fits to
the measured values.  Contaminants which were produced in the target
and could be transported as gas were $^{11}$C ($t_{1/2} =
20.34$~m), $^{13}$N ($t_{1/2} = 9.96$~m), and $^{15}$O ($t_{1/2}
= 122.2$~s) via the reactions $^{12}$C($ ^3$He,$\alpha$)$^{11}$C,
$^{12}$C($^3$He,d)$^{13}$N, and $^{13}$C($^3$He,n)$^{15}$O.  The mass
resolution of the separating magnet in the ion beam line was $\delta
M/M = 0.53\%$, and ion beam contamination at the target was estimated
to be less than one part per million of the next charge-to-mass ratio
species.  However, these $\beta^{+}$ emitters could be transported as
neutral gas through the cryogenic trap at the entrance to the ion
source, through the source (but not ionized), and into the counting
chamber by molecular diffusion.  We allowed the amounts of these
activities to vary in the fits to the decay data, finding amounts of
contaminant activity of $10^{-5}$ to $10^{-3}$ of the $^{14}$O
activity (depending on the time bin in which the fits were started).  
The count rate in the fit function was calculated for the
counting time of 0.4995~s and corrected for dead time loss to obtain a
theoretical decay curve.  The spacing of the time bins in the fits was
0.5~s, corresponding to the total time spent in each bin in the
experiment.  This theoretical decay curve was compared to the data to
generate the maximum likelihood estimator in the fits.  The data for
each of three detectors (labeled B, C, and D) and three dead times
were analyzed separately for each run.  The final average was obtained
from the unweighted mean of the half-life results from all 29 runs.
We used an unweighted mean since the statistical uncertainty in the
$^{14}$O $t_{1/2}$ determined by the fit was nearly identical for
each run at a given initial count rate, while the variation in the
fitted half-life exceeded the statistical uncertainty found by the
fit.  We varied the initial count rate (the ``start time'' of the fit)
in the analysis to search for systematic errors.  Detector (A) failed
during the run when its high-voltage lead shorted.  We observed
intermittent data loss and high-voltage discharge for this detector in
several runs, and we did not use this detector in the final analysis.
A typical decay curve for a single run in one detector is shown in
Fig.~\ref{fig:decaycycle}.  The statistical uncertainty in the
$^{14}$O half-life from a fit to a single run with an initial count
rate of 20~kHz was about 65~ms.

Using a Monte-Carlo simulation, we investigated several potential
systematic effects caused by instability in the detectors and
electronics.  We produced simulated decay data, including dead time
losses.  To study the effect of drifts, we introduced a linear
time-dependent detector efficiency into the simulation.  Simulated
data were fit to determine the dependence of the half-life on the
drift.  We measured drifts in time and with temperature of the PMT
high voltage power supplies, discriminator voltage set points, and PMT
gains.  In off-line tests using a radioactive $^{90}$Sr beta source,
we measured the count rate shifts induced by these drifts, and then
interpolated the results from the Monte-Carlo data to estimate a
systematic uncertainty in the half-life.  We also studied the effect of
dead time uncertainty with this technique, finding a 2~ms uncertainty
in the half-life (from fits at 20~kHz initial rate) caused by the
uncertainty of 9~ns in the measured dead time.  This agrees with the
2-5 ms differences observed in the averaged half-life measurements for
the three different dead time channels.  We neglect the effect of
short, earlier dead times in series with the long dead time logic
pulses sent to the scalers.  This would cause an error of less than
1~ms in the measured half-life using data with an initial rate of
20~kHz.  We observed fluctuations of 20-50\% (with a period of about
45 s) in the average background rate in the detectors both during and
after the runs.  We modelled this behavior in Monte-Carlo generated
data to determine the effect on the half-life, and assigned systematic
uncertainties to each detector of 2~ms (B), 9~ms (C), and 3~ms (D).
The average background rates were 0.106~Hz, 0.558~Hz, and 0.189~Hz,
respectivly.  The fluctuation of the PMT's high voltage (of about 0.3
volts) causes an uncertainty of 2 ms in the half-life for the
data with 20 kHz initial count rate.

\begin{figure}
\includegraphics[width=0.5\textwidth]{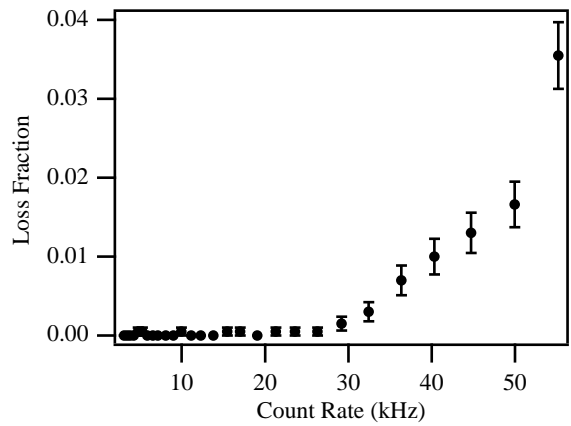}
\caption{\label{fig:rateshift} 
The fraction of counts in detector D lost below threshold 
as a function of the count rate in the detector.}
\end{figure}

We measured gain changes in the detectors as a function of the count
rate.  Before taking the data to measure the $^{14}$O half-life, we
measured the pulse height spectra of the scintillators during a
bombardment and counting cycle with the same initial count rates in
the scintillators (up to 180~kHz in the front detectors) as in the
half-life measurement data.  The data for this run do not have a
precise time base and were not used to determine the half-life.  The
most probable pulse height in the counters was smaller at high count
rates than at lower rates.  This caused a fraction of the counts to be
lost below threshold at high rates.  Figure~\ref{fig:rateshift} shows
this count loss fraction as a function of rate.  In the worst case
(detector A), we estimate that 15\% of the counts in the
minimum-ionizing spectrum in the detector can be shifted below
threshold at the highest rates observed during the experiment
(200~kHz).  We also performed an off-line test of the detectors to
search for transient time behaviors of the gains.  We used a $^{90}$Sr
beta source placed behind a movable shutter to produce a rapid rate
change in the detectors (from less than 1~Hz to 12~kHz in less than
0.1 seconds).  We observed a small, transient change in the count rate
in one of the detectors immediately after the change to high rate.
The rate in detector D increased by 0.6\% to its steady-state value in
about 125~seconds.  This transient effect was not present above 0.05\%
in a second detector we tested (detector C).  To avoid systematic
error in the half-life caused by the change in gain in the detectors
as a function of time or rate, we restricted the analyzed data to
count rates less than 20~kHz.  This removed approximately the first
150 seconds of data in each run.  In Fig.\ ~\ref{fig:rateshift}, this
cut reduces the measured count loss fraction to less than $5\times
10^{-4}$ of the total counts per bin.  At this level, according to the
simulations, the count loss from any remaining gain shift would change
the half-life by less than 4~ms.

\begin{figure}
\includegraphics[width=0.5\textwidth]{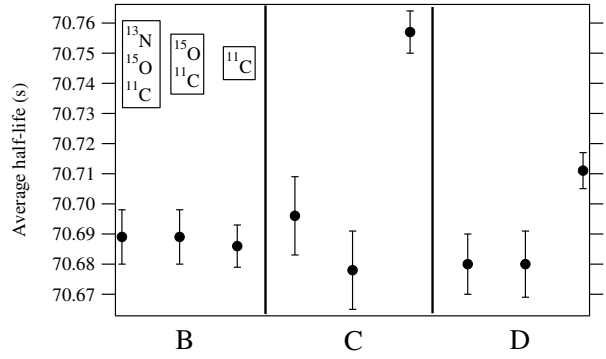}
\caption{\label{fig:contam} 
The half-life results for fits to data from each detector when
allowing different combinations of contaminant species to vary in the
fits.  Each point is an unweighted average of the half-life
results from the 29 data cycles in which the data is restricted to an
initial detector count rate of 20 kHz.}
\end{figure}

The largest sytematic uncertainty in our result is caused by the
presence of contaminant $\beta$ decay activities.  We performed
several analyses which included different combinations of the possible
contaminant species $^{11}$C, $^{13}$N, and $^{15}$O as fit terms.
The amount of contaminant species identified by the fits ranged from
$10^{-5}$ of the initial $^{14}$O activity (for $^{11}$C) to $10^{-3}$
(for $^{15}$O when using a starting time bin for the fits
corresponding to count rates of 20~kHz in the detectors).  When
starting the fits at different time bins to limit the initial count
rates in the detectors and to search for other rate-dependent effects,
the amount of contaminant activity (extrapolated to the beginning of
the counting cycle in the run) was consistent for each species as we
varied the start time of the fits.  The fraction of contaminant
species varied among runs by about a factor of two, and the fraction
of contaminant species identified by the fit routines agreed among the
detectors in each run.  We found no statistically significant
difference in goodness-of-fit for fits using all three contaminants,
fits using only $^{11}$C, and fits using $^{11}$C and $^{15}$O.  The
resulting averaged $^{14}$O half-life for these three different fit
methods differed by 3~ms for detector B, 53~ms for detector C, and
13~ms for detector D, when the fits were restricted to data with a
20~kHz initial rate.  This suggests that we should assign a systematic
uncertainty to our inability to distinguish which of the contaminant
contributions is the best fit.  The averaged half-life for each
detector is shown in Figure~\ref{fig:contam} for each of three
analyses with different combinations of contaminant species allowed to
vary in the fits.  Detector B has no disagreement among the fits,
while detector C shows a strong disagreement for the result of the fit
with only $^{11}$C.  This may be due to the relatively higher
background rate in detector C (0.55~counts per second) compared to
detector B (0.11~counts per second), as well as the fact that to
achieve 20~kHz initial rate in C, we have to wait longer, so that the
ratio of $^{14}$O activity to contaminant is necessarily lower.  Our
simulations (in which we artificially introduce contaminant species
into the data, while not allowing these actvities to vary in the fits)
bear out this explanation.  To assign a systematic uncertainty, we
take an unweighted average over the three fit methods with different
contaminant combinations for each detector.  For each detector, we
could use the standard deviation of this mean as an uncertainty
associated with our inability to fully determine the contaminant
identities using only the fits to the decay data. This would be a 2~ms
uncertainty for detector B, while detector C would have a 41~ms
uncertainty, and an 18~ms uncertainty for detector D.  In combining
the data from different detectors for a final average, however, we
believe that this would underestimate the contaminant identification
uncertainty, since a final systematic uncertainty estimate would be
dominated by the estimate in detector B.  Instead, to estimate the
systematic uncertainty from contaminant ambiguity, we use the average
run-to-run difference in half-life among the three fit methods for
each detector: 47~ms (detector B), 54~ms (detector C), and 55~ms
(detector D).  In general, restricting the analysis to low initial
count rates to avoid the errors caused by gain shifts in the detectors
forces accepting a large uncertainty from being unable to identify the
contaminating beta activities (particularly $^{15}$O).  For the final
result, we use the half-life values from fits which allowed the
amounts of all three possible contaminant species to vary.

The measured half-life for each detector is derived from an unweighted
average of the fit results from the 29 runs.  In this average, we
arbitrarily select one of the dead time channels (since the results
from the three dead time channels agree within 5~ms in this analysis and
since the data in each channel are not statistically independent).  We
use an unweighted average since the chi-squared for a weighted average
(with the half-life uncertainty for each run equal to the statistical
error from the fit) over the 29 runs is high ($\chi^{2}/\text{d.o.f.}
\approx 2.5$).  The large scatter among the run results is probably
caused by variation of the amount and identity of contaminant activity
and constant background rate (or its drift) from run to run.  The
statistical uncertainties in the unweighted averages (the standard
deviation of the mean of the 29 runs) are 11~ms (detector B), 17~ms
(C), and 11~ms (D).  We apply the systematic uncertainties (added in
quadrature) to each detector from deadtime uncertainty (4~ms); drifts
in background and voltage (B, 2.5~ms; C, 9~ms; D, 4~ms); and
contaminant identification (as above), and then average over the
detectors.  The result with statistical and systematic uncertainty for
each detector are B: 70.698(11)(47) seconds, C: 70.697(17)(55)~s, and
D: 70.688(11)(55)~s.  We cannot average over all three detectors,
because detectors C and D (front and back) count many of the same
$\beta$ particles (which can trigger minimum ionizing pulses in both
detetors) and are not statistically independent.  We can only average
detectors B and C or detectors B and D.  To generate a final
uncertainty, we reduce the statistical uncertainty component, while
using the mean systematic uncertainty of the detectors.  Averaging B
and C yields 70.698(9)(51), while averaging B and D yields
70.693(8)(51).  Our result is a mean of these two values (without
reducing uncertainty), $t_{1/2}[^{14}\text{O}] = $70.696(52)~s, with
statistical and systematic uncertainties added in quadrature.

Our technique using a mass-separated radioactive beam produced samples
with relatively low contamination and low background.  Our result
could be improved with the use of stabilized PMT counter systems to
decrease the potentially large systematic error from changing gains.
Avalanche photodiodes generally have a wider dynamic range in rate
than PMTs and also seem promising.  These approaches would allow an
experiment to take advantage of the small statistical error when using
fast counters and an intense source of activity.  With stabilized
gains, we could count more decades of decay activity and better
resolve the contaminant contributions to the decay data.

Our result for $t_{1/2}[^{14}\text{O}]$ is $1.5\sigma$ longer than the
recommended average value 70.616(14) in Ref.\ ~\cite{hardy04}.  The
authors of Reference \cite{barker04}, who found
$t_{1/2}[^{14}\text{O}] = 70.641(20)$, suggested that earlier
half-life measurements (e.g. Clark {\it et al.} \cite{clark73}) were
subject to error from pileup effects in precise measurements performed
with germanium detectors. Our result with faster scintillators and
logic electronics supports this hypothesis.  If we include our
measured half-life in the average calculated by Towner and Hardy in
Ref.\ ~\cite{hardy04}, we obtain an average half-life of 70.620(14)
(following the prescription to renormalize the uncertainty when
$\chi^{2}/\text{d.o.f.} > 1$).  Selecting a different set of
measurements (on the basis of avoiding rate-dependence or pileup
error) to derive a ``best value'' for the half-life of $^{14}$O could
arrive at a higher value.  Using the average half-life value
70.620(14) and the rest of the data (including the recent measurement
of the $^{14}$O Q-value by Tolich {\it et al.}  in Ref.\
~\cite{tolich03}), the re-evaluated branching ratio in Ref.\
~\cite{towner05}, the calculated corrections $\delta_{R}$,
$\delta_{C}$, and the re-calculated electron capture probability for
$^{14}$O summarized in Ref.\ \cite{hardy04}, we obtain ${\cal
F}t(^{14}\text{O}) = 3068.2(2.6)$.  This is to be compared with the
values calculated in Ref.\ \cite{savard05} for $^{14}$O of ${\cal
F}t(^{14}\text{O}) = 3072.0(2.6)$, and to the average ${\cal F}t$
value from the twelve most precisely measured superallowed
transitions, $\left\langle {\cal F}t \right\rangle = 3074.4(1.2)$.  A
recalculation of the unitarity sum is presented in \cite{savard05},
using a value for $V_{us}$ suggested in \cite{mescia04}.

\begin{acknowledgments}
We thank the staff at the 88-Inch Cyclotron for their assistance
during the project.  S.W. Leman worked on early phases of the
experiment and N.D. Scielzo assisted in data acquisition.  This work
was supported by the Director, Office of Science, Office of Basic
Energy Sciences, of the U.S. Department of Energy under Contract
No. DE-AC02-05CH11231.
\end{acknowledgments}


\end{document}